\RequirePackage{lineno}

\documentclass[aps,twocolumn,prb,amsmath,amssymb,superscriptaddress,floats]{revtex4}

\usepackage{amsmath}	
\usepackage{gensymb}
\usepackage{hyperref}
\hypersetup{colorlinks=true}
\usepackage{upgreek}
\usepackage{graphics}
\usepackage{epsfig}
\usepackage{color}
\usepackage{bm}
\usepackage{float}
\usepackage[normalem]{ulem}     
\usepackage{dsfont}             
\usepackage{graphicx}
\usepackage{marvosym}           

\graphicspath{{./}{./plots/}}

\newif\ifjustafterhead
\newcommand{\artsec}[1]{\section*{}\vspace{-1.00cm}\noindent{\large\textsf{\textbf{#1}}}\par\vspace{0.12cm}\justafterheadtrue}
\newcommand{\artsub}[1]{\ifjustafterhead\vspace{0.06cm}\else\vspace{0.50\baselineskip}\fi\justafterheadfalse\noindent{\textbf{#1}}}

\newcommand{\corrtag}{\textsuperscript{\Letter}}

\newcounter{smnote}
\newcommand{\smnote}[2]{\par\vspace{0.55cm}{\centering\refstepcounter{smnote}\label{#2}\textsf{\textbf{Supplementary Note \thesmnote: #1}}\par}\vspace{0.28cm}}

\makeatletter
\newbox\bfig@box
\newdimen\bfig@ht
\newdimen\bfig@sep \bfig@sep\z@
\newif\ifbfig@queued
\newif\ifbfig@active

\let\bfig@orig@set@colht\set@colht
\def\set@colht{%
  \bfig@orig@set@colht
  \ifbfig@queued
    \global\bfig@queuedfalse
    \global\bfig@activetrue
  \fi
  \ifbfig@active
    \global\advance\@colht by -\dimexpr\bfig@ht+\bfig@sep\relax
    \global\advance\@colroom by -\dimexpr\bfig@ht+\bfig@sep\relax
  \fi
}

\def\bfig@ship{%
  \ifbfig@active
    \setbox\@outputbox\vbox to\textheight{%
      \unvbox\@outputbox
      \vfil
      \vskip\bfig@sep
      \box\bfig@box
    }%
    \global\bfig@activefalse
    \global\setbox\bfig@box\box\voidb@x
  \fi
}
\prepdef\@outputpage{\bfig@ship}

\newenvironment{widebottomfigure}{%
  \par
  \refstepcounter{figure}%
  \global\setbox\bfig@box\vbox\bgroup
    \hsize\textwidth \linewidth\hsize \columnwidth\hsize
    \centering
}{%
  \egroup
  \global\bfig@ht=\dimexpr\ht\bfig@box+\dp\bfig@box\relax
  \global\bfig@queuedtrue
  \par
}
\AtBeginDocument{\bfig@sep\dbltextfloatsep\relax}
\newcommand{\bfigcaption}[1]{\@makecaption{\fnum@figure}{#1}}
\makeatother

\begin{document}
\justafterheadfalse


\title{Emergent $\mathrm{SU}(2)_2$ Wess--Zumino--Witten criticality\\ in an anisotropic spin-1 Kitaev--$\Gamma$ chain}

\author{Qiang Luo\corrtag}
\affiliation{College of Physics, Nanjing University of Aeronautics and Astronautics, Nanjing 211106, China}
\affiliation{Key Laboratory of Aerospace Information Perception and Physics (NUAA), Nanjing 211106, China}

\author{Shijie Hu}
\affiliation{Beijing Computational Science Research Center, Beijing 100084, China}

\author{Wang Yang\corrtag}
\affiliation{School of Physics, Nankai University, Tianjin 300071, China}

\author{Yuhai Liu}
\affiliation{School of Physical Science and Technology, Beijing University of Posts and Telecommunications, Beijing 100876, China}

\author{Jinbin Li}
\affiliation{College of Physics, Nanjing University of Aeronautics and Astronautics, Nanjing 211106, China}
\affiliation{Key Laboratory of Aerospace Information Perception and Physics (NUAA), Nanjing 211106, China}

\author{Jize Zhao}
\affiliation{School of Physical Science and Technology $\&$ Lanzhou Center for Theoretical Physics, Lanzhou University, Lanzhou 730000, China}

\author{Xiaoqun Wang\corrtag}
\affiliation{School of Physics, Zhejiang University, Hangzhou 310058, China}

\date{\today}

\begin{abstract}
Bond-directional interactions underpin unconventional quantum
phenomena, yet their interplay with single-ion anisotropy
remains elusive. Here we combine large-scale density-matrix
renormalization group calculations and bosonization to
investigate the spin-1 Kitaev--$\Gamma$ chain.
Positive anisotropy drives a smooth crossover from the
Kitaev regime to the trivial large-$D$ limit, contrasting
with the topological Haldane-to-large-$D$ transition in
the spin-1 Heisenberg chain. This adiabatic evolution
reshapes the excitation spectrum, nonlocal correlations,
and thermodynamic signatures. Conversely, negative
anisotropy induces a changeover from a first-order to
a continuous dimer--Haldane transition. The continuous
transition exhibits $\mathrm{SU}(2)_2$ Wess--Zumino--Witten
criticality with central charge of $3/2$, despite
the absence of microscopic continuous symmetry.
Bosonization predicts a spin-correlation exponent
$\eta = 3/4$, consistent with our numerical estimate
of $0.77(3)$. Our findings reveal how bond-directional
interactions shape distinctive gapped regimes and
emergent conformal criticality.
\end{abstract}


\pacs{}

\maketitle

{\let\thefootnote\relax\footnotetext{\Letter~e-mail:
qiangluo@nuaa.edu.cn (Q.L.); wyang@nankai.edu.cn (W.Y.);
xiaoqunwang@zju.edu.cn (X.W.)}}


\artsec{Introduction}

Quantum spin chains provide a setting for understanding how topology and symmetry
shape quantum phase transitions~\cite{Kosterlitz1973,Senthil2004Sci,Balents2010Nat,Matsuda2025RMP,Belitz2017PRL,Zhao2026PRX}.
The spin-1 Heisenberg chain hosts the Haldane
phase~\cite{Haldane1983PRL}, a paradigmatic symmetry-protected topological
phase~\cite{Chen2012Sci,Pollmann2010PRB} recently realized in metal-organic
frameworks~\cite{Tin2023NC}. A uniaxial
single-ion anisotropy (SIA) $D$ can drive this state into a topologically trivial
large-$D$ phase through a quantum phase
transition~\cite{Chen2003PRB,Hu2011PRB,Tzeng2017PRB}. Beyond conventional phase
boundaries, a \textit{crossover} describes a smooth evolution between two regimes
without a thermodynamic singularity~\cite{SaDeMelo2024AR}, while a
\textit{changeover} denotes the transformation of a first-order transition into a
continuous one upon tuning a control
parameter~\cite{Jin2018PRB,Ejima2018SPP,Jian2021PRL,Chepiga2019PRL}. Spin-1 chains with
additional interactions realize this paradigm: the dimer--Haldane transition
belongs to the $\mathrm{SU}(2)_2$ Wess--Zumino--Witten (WZW) universality class
for weak frustration but becomes first-order as the couplings
vary~\cite{Pixley2014PRB,Chepiga2016PRBb,Chepiga2016PRBa,Chepiga2016PRBc}. These
results motivate the question of how strongly anisotropic interactions modify the
connection between topology, symmetry, and criticality.

Bond-directional interactions offer a route to addressing this question because
they couple spin components to lattice directions and generally remove continuous
spin-rotation symmetry. In two dimensions, this paradigm is exemplified by the
exactly solvable Kitaev honeycomb model, which hosts a quantum spin-liquid ground
state~\cite{Kitaev2006} whose magnetic interactions can be tuned in
situ~\cite{Kumar2022CP}. A symmetry-allowed $\Gamma$ term further enriches the
phase diagram~\cite{Ran2014PRL}, giving rise to proximate and $\Gamma$-type spin
liquids~\cite{Wang2019PRL,Luo2021npj}. Reducing dimensionality from this
model, bond-directional spin chains have emerged as a fertile ground for exotic
quantum
phenomena~\cite{Agrapidis2018SR,Yang2020PRL,Yang2021PRB,Luo2021PRB,Luo2021PRR,Sorensen2021PRX,Macedo2022PRB,Sorensen2023PRRa,Sorensen2023PRRb,Luo2023PRB,Saito2024PRL,Wei2025PRB,Reja2025arXiv,Bonkhoff2025PRL,Raja2026PRL},
including emergent symmetries~\cite{Yang2020PRL}, deconfined
criticality~\cite{Macedo2022PRB,Luo2023PRB}, and exact solvability in certain
limits~\cite{Saito2024PRL,Reja2025arXiv,Raja2026PRL}. Notably, the spin-$1/2$
Kitaev--$\Gamma$ chain supports an emergent low-energy SU(2) symmetry despite its
spin anisotropy~\cite{Yang2020PRL}. In the spin-1 counterpart, Kitaev
interactions instead produce a gapped nonmagnetic regime with
spin-nematic correlations and a double-peak specific
heat~\cite{Luo2021PRR,Luo2023PRB,Jiang2023PRL}. Whether a transition involving the Haldane
phase can realize level-2 WZW criticality in this bond-directional setting, and
how single-ion anisotropy controls it, remain unresolved.

Single-ion anisotropy---a decisive ingredient in spin-1
magnets~\cite{Do2023npj}---introduces two contrasting possibilities in the spin-1
Kitaev chain. For positive $D$, previous work has established a smooth, gapped connection
to the large-$D$ limit~\cite{Sorensen2023PRRb}; the evolution of excitations,
nonlocal correlations, and thermodynamic responses nevertheless lacks a unified
characterization. For negative $D$, a spontaneously dimerized phase
emerges~\cite{Luo2023PRB}, allowing its competition with the Haldane phase upon
introducing $\Gamma$ exchange. The central questions are whether anisotropy
changes the order of the dimer--Haldane transition and which critical theory
describes its continuous segment without microscopic continuous spin-rotation
symmetry.

\begin{figure}[tbp]
  \centering
  \includegraphics[width=0.95\columnwidth, clip]{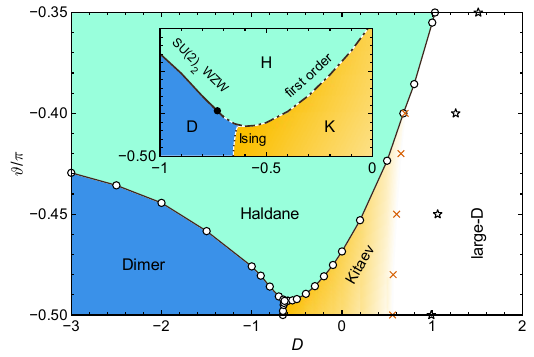}
  \caption{\textbf{Quantum phase diagram of the spin-1 Kitaev--$\Gamma$ chain.} The ground-state phases comprise the dimerized, Haldane,
  Kitaev, and large-$D$ regimes. Open circles denote phase boundaries; crosses
  ($\times$) (from order-parameter intersections) and pentagrams ($\bigstar$)
  (from specific-heat double peaks) mark crossovers between the Kitaev and
  large-$D$ regimes. Inset, magnified view of the region $-1\le D\le0$; line
  styles (dotted, dotted-dashed, solid) distinguish Ising, first-order, and
  $\mathrm{SU}(2)_2$ WZW transitions. The black dot marks the changeover between
  the first-order and continuous segments at $(D_c,\vartheta_c/\pi) \approx (-0.73,-0.4893)$.}
  \label{fig:phase}
\end{figure}

Here we combine zero-temperature~\cite{White1992,Schollwock2005} and
finite-temperature~\cite{BurXiangGeh1996,WangXiang1997} density-matrix
renormalization group (DMRG) calculations with bosonization
analysis~\cite{Gogolin1998,Giamarchi2004} to address these questions in the
spin-1 Kitaev--$\Gamma$ chain (Fig.~\ref{fig:phase}). Positive anisotropy
produces a crossover towards the large-$D$ limit, with smooth changes in the
excitation gap and nonlocal correlations and the disappearance of the
low-temperature specific-heat peak. For negative anisotropy, increasing $|D|$
changes the dimer--Haldane transition from first order to continuous. At a
representative continuous transition, we obtain $c=1.51(2)$ and
$\eta=0.77(3)$, consistent with the $\mathrm{SU}(2)_2$
WZW predictions $c=3/2$ and $\eta=3/4$. Together with the low-energy
field-theory analysis, these results support emergent level-2 WZW criticality in
a spin-1 chain without microscopic continuous spin-rotation symmetry and
establish the Kitaev chain as a minimal model for tracking the route from
crossover to changeover.


\artsec{Results}

\artsub{Model and phase diagram}

The Hamiltonian of the spin-1 Kitaev--$\Gamma$ chain with a uniaxial SIA is given
by
\begin{align}\label{eq:ham}
\mathcal{H} = \sum_{l=1}^{L/2} \left[\mathcal{H}_{2l-1,2l}^{(x)}(\vartheta) + \mathcal{H}_{2l,2l+1}^{(y)}(\vartheta)\right] + D \sum_{l=1}^{L} (S_l^z)^2 ,
\end{align}
where $L$ is the chain length. The exchange part on a bond of type
$\gamma\in\{x,y\}$, with $\{\alpha,\beta,\gamma\}$ cyclic, reads
$\mathcal{H}_{i,j}^{(\gamma)}(\vartheta) = K S_i^{\gamma}S_j^{\gamma} +
\Gamma (S_i^{\alpha}S_j^{\beta}+S_i^{\beta}S_j^{\alpha})$, where $K=\sin\vartheta$
and $\Gamma=\cos\vartheta$. Following a $U_6$ rotation that eliminates the cross
terms~\cite{Yang2020PRL}, the exchange part becomes
$\tilde{\mathcal{H}}_{i,j}^{(\gamma)}(\vartheta) = -K \tilde{S}_i^{\gamma} \tilde{S}_j^{\gamma}
+ \Gamma (\tilde{S}_i^{\alpha}\tilde{S}_j^{\alpha}+\tilde{S}_i^{\beta}\tilde{S}_j^{\beta})$
on bonds $\gamma=\tilde{x},\tilde{y},\tilde{z}$ circularly. The same rotation
recasts the SIA term as
$\tilde{\mathcal{H}}_{D} = D \sum_{l=1}^{L/3} [(\tilde{S}_{3l-2}^z)^2 +
(\tilde{S}_{3l-1}^y)^2 + (\tilde{S}_{3l}^x)^2]$, revealing a three-site
periodicity in the rotated basis. Our phase diagram is determined by DMRG
calculations, in which both periodic (PBC) and open (OBC) boundary conditions are
employed depending on the observables.

Figure~\ref{fig:phase} summarizes the ground-state phases in the parameter range
$\vartheta/\pi\in[-0.50,-0.35]$ and $D\in[-3,2]$. In the absence of SIA, the
system is in the gapped Kitaev phase, characterized by long-range spin-nematic
correlations and a double-peak specific
heat~\cite{Luo2021PRR,Luo2023PRB}. Turning on the $\Gamma$ interaction drives a
first-order transition to the Haldane phase at
$\vartheta_t/\pi=-0.4685(5)$~\cite{Luo2021PRR}. For positive $D$, the ground
state is the trivial large-$D$ phase. Although no phase transition separates the
Kitaev and large-$D$ phases~\cite{Sorensen2023PRRb}, a crossover between them is
identified by the evolution of the excitation gap, nonlocal order parameters, and
the specific heat (detailed below). For negative $D$, a spontaneously dimerized
phase emerges via an Ising transition from the Kitaev
phase~\cite{Luo2023PRB}. The transition from the dimerized to the Haldane phase
changes character from first order to continuous (with central charge $3/2$,
belonging to the $\mathrm{SU}(2)_2$ WZW universality class) as $|D|$ increases.

\artsub{Crossover from Kitaev to large-$D$ regime}

\begin{figure*}[tbp]
\centering
\includegraphics[width=0.95\textwidth, clip]{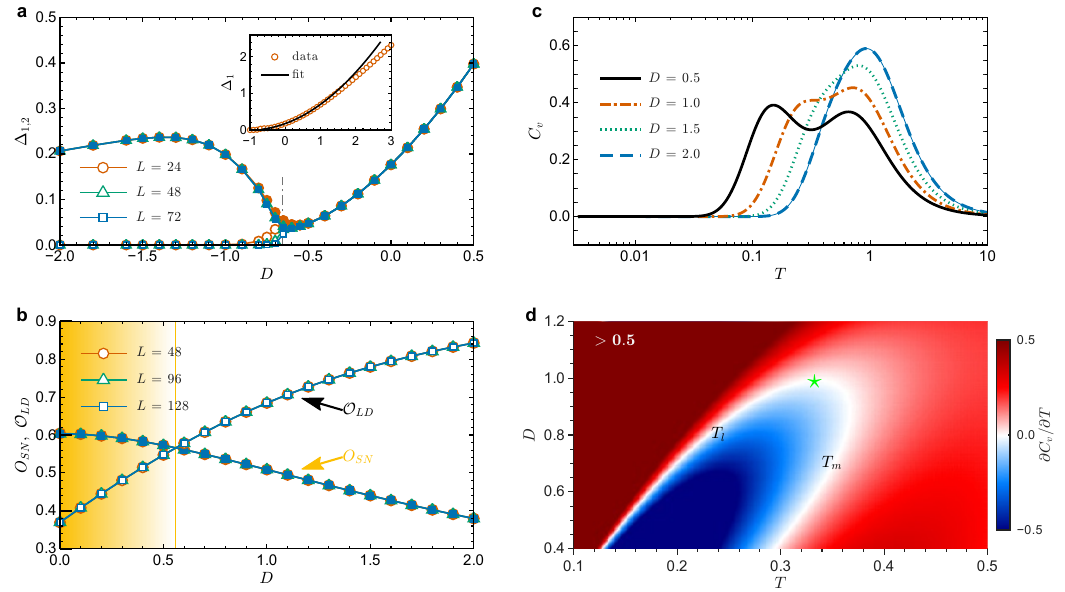}
\caption{\textbf{Kitaev-to-large-$D$ crossover under positive SIA.}
\textsf{\textbf{a}}, Excitation gaps $\Delta_{1,2}$ as a function of $D$ at
$\vartheta/\pi=-1/2$ for different chain lengths $L=24$ (red
circles), $48$ (green triangles), and $72$ (blue squares) under PBC. Inset,
quadratic fit of $\Delta_1$ at small $D$, where the deviation from the
quadratic form signals the onset of the crossover. \textsf{\textbf{b}},
Spin-nematic order parameter $O_{SN}$ (filled symbols) and string order
parameter $\mathcal{O}_{LD}$ (open symbols), characterizing the Kitaev and
large-$D$ phases, respectively, as a function of $D$ for different chain lengths
$L=48$ (red circles), $96$ (green triangles), and $128$ (blue squares) under
OBC. \textsf{\textbf{c}}, Specific heat $C_v$ as a function of temperature $T$
for selected SIA parameters $D=0.5$ (solid line), $1.0$ (red), $1.5$ (green),
and $2.0$ (blue). \textsf{\textbf{d}}, Color map of $\partial C_v/\partial T$ in
the $(T,D)$ plane. The curves $T_l$ and $T_m$ indicate the loci where
$\partial C_v/\partial T=0$, with the star marking where the low-$T$ peak
disappears.}
\label{fig:crossover}
\end{figure*}

For the ferromagnetic Kitaev coupling ($\vartheta=-\pi/2$), positive SIA does not
destabilize the Kitaev phase through a phase transition; instead, the system
crosses over smoothly to the conventional large-$D$ phase. The only boundary at
negative SIA is an Ising quantum phase transition to a dimerized phase at
$D_c=-0.6551(2)$~\cite{Luo2023PRB} (Supplementary Fig.~\ref{fig:SMfss}); for
$D>D_c$, the ground state is unique and no further transition occurs. This crossover is first evidenced by the lowest excitation gap
$\Delta_1$: near the pure Kitaev point ($D=0$),
$\Delta_1(D) \simeq \Delta_K (D+1)^2$ with $\Delta_K=0.17634970(2)$, reflecting
the quadratic dispersion of unusual excitations in the Kitaev
phase~\cite{Luo2021PRR}; in the large-$D$ limit, $\Delta_1\propto D$, as expected
for a gapped paramagnet with single-magnon excitations. The deviation from the
quadratic form becomes appreciable for $D\gtrsim1$ [inset of
Fig.~\ref{fig:crossover}a], signaling the onset of the crossover.

We further characterize the two disordered phases using distinct order parameters.
The Kitaev phase supports a long-range four-spin correlation
$Q_1(i,j) = \langle S_i^+ S_{i+1}^+ S_j^- S_{j+1}^- \rangle \simeq
{O}_{SN}^2 e^{-i\phi}$, from which we extract the spin-nematic order parameter
${O}_{SN}$~\cite{Luo2023PRB}. The large-$D$ phase is instead captured by a
parity-like string order parameter
$\mathcal{O}_{LD} = - \lim_{|q-p|\to\infty} \langle \prod_{r=p}^q e^{i\pi S_r^z}
\rangle$~\cite{denNijsRom1989PRB}. As shown in
Fig.~\ref{fig:crossover}b, both order parameters vary smoothly with $D$ and
cross each other at $D \approx 0.55$. However, their derivatives exhibit clear
anomalies: the first derivative of ${O}_{SN}$ displays a broad dip near
$D\approx0.85$, and the second derivative of $\mathcal{O}_{LD}$ shows a local
minimum around $D\approx0.75$ (not shown). The smoothness of the order parameters
themselves, contrasted with the extrema in their derivatives, is a hallmark of a
crossover rather than a phase transition, placing the crossover scale at
$D\sim0.8$.

Thermodynamic quantities provide independent confirmation. In the pure Kitaev
limit, the specific heat $C_v$ exhibits two distinct peaks at $T_l\simeq0.0562$
and $T_h\simeq0.5680$, with a valley at $T_m\simeq0.1991$---a fingerprint of
multiple excitations~\cite{Luo2021PRR}. As $D$ increases, the low-temperature
peak gradually weakens and vanishes for $D\gtrsim1$
[Fig.~\ref{fig:crossover}c]. To pinpoint this disappearance, we examine the
contour of $\partial C_v/\partial T$ [Fig.~\ref{fig:crossover}d]: the two
low-temperature zeros, corresponding to $T_l$ and $T_m$, approach each other and
merge at $D\approx0.98(1)$, defining the threshold above which the double-peak
structure is lost. Together with the gap deviation and the order-parameter
anomalies, three independent probes thus converge on the crossover scale
$D\sim0.8$--$1.0$, reinforcing the conclusion that the system crosses over from
the Kitaev phase to a large-$D$ phase. The same suppression of the double-peak structure is found at
$\vartheta/\pi=-0.45$, $-0.40$, and $-0.35$ (Supplementary Fig.~\ref{fig:SMcv}), so that the
crossover is a property of the bond-directional chain rather than of the Kitaev
point alone.

\begin{figure}[htbp]
\centering
\includegraphics[width=0.95\columnwidth, clip]{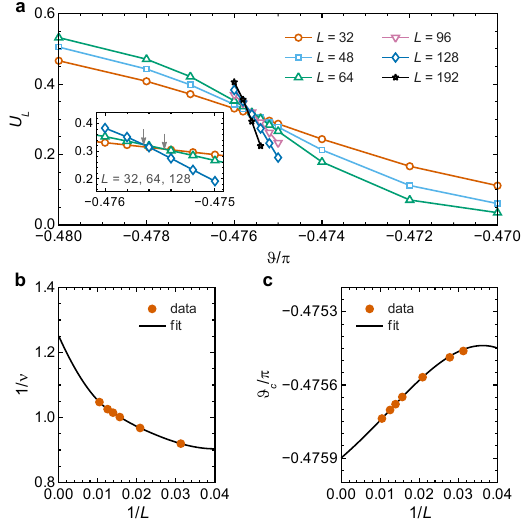}
\caption{\textbf{Binder-cumulant analysis of the dimerized-to-Haldane
transition.} \textsf{\textbf{a}}, Binder cumulant $U_L$ for different chain
lengths $L$ ranging from 32 to 192 in the Kitaev--$\Gamma$ spin chain with SIA
$D=-1.0$. Inset, zoom-in near the quantum critical point for chain lengths
$L=32$ (red circle), $L=64$ (green triangle), and $L=128$ (blue diamond).
\textsf{\textbf{b}},\textsf{\textbf{c}}, Estimates of the inverse critical exponent
$1/\nu$ [cf. Eq.~\eqref{eq:nu}] and of the critical point $\vartheta_{c}$,
obtained from pairs of chain lengths $(L, 2L)$.}
  \label{fig:binder}
\end{figure}

\begin{widebottomfigure}
\includegraphics[width=0.95\textwidth, clip]{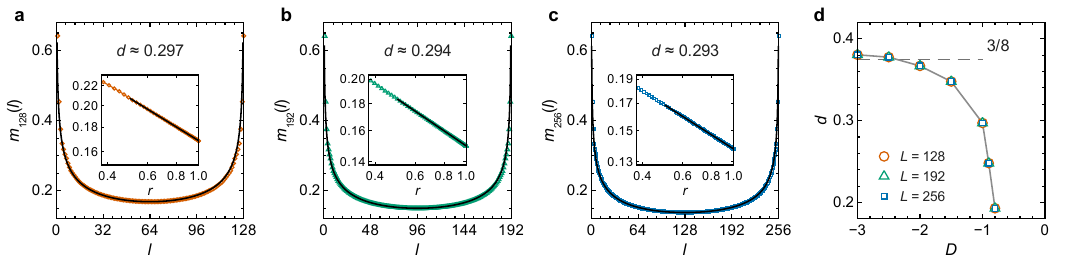}
\bfigcaption{\textbf{Scaling dimension of the local dimerization.}
\textsf{\textbf{a}}--\textsf{\textbf{c}}, Site dependence of the local dimerization
$m_{L}(l)$ in the Kitaev--$\Gamma$ spin chain with SIA $D=-1.0$ for chain lengths
\textsf{\textbf{a}}, $L=128$ (red circle); \textsf{\textbf{b}}, $L=192$ (green triangle);
and \textsf{\textbf{c}}, $L=256$ (blue square). The inset in each panel shows the
extraction of the scaling dimension $d$. \textsf{\textbf{d}}, Scaling dimension $d$
along the critical line as a function of $D$; the horizontal dashed line marks
the exact $\mathrm{SU}(2)_2$ value $d=3/8$.}
\label{fig:dimension}
\end{widebottomfigure}

\artsub{Changeover from first-order to continuous transition}

To uncover the nature of the dimer--Haldane transition at negative SIA, 
we measure the order parameters of the corresponding phases. 
The spontaneously dimerized phase has a twofold-degenerate ground state in the thermodynamic limit. 
We define the local dimerization operator as 
$\hat m_{2l-1}= S_{2l-1}^{x}S_{2l}^{x}-S_{2l-2}^{y}S_{2l-1}^{y}$ and $\hat m_{2l}=S_{2l-1}^{x}S_{2l}^{x}-S_{2l}^{y}S_{2l+1}^{y}$, 
whose expectation value $m_L(l)=\langle\hat m_l\rangle$ measures the difference 
between the nearest-neighbor spin correlations on adjacent $x$ and $y$ bonds.
The dimer order parameter is $M_L=\langle\hat M_L\rangle$, 
where $\hat M_L=(1/L)\sum_{l=1}^{L}\hat m_l$ is the spatially averaged dimerization operator. 
In comparison, the Haldane phase is a symmetry protected topological phase 
and is characterized by the nonlocal string order parameter
$\mathcal{O}_H = - \lim_{|q-p|\to\infty} \Big\langle \tilde{S}_p^{z}
\big(\prod\limits_{p<r<q} e^{i\pi \tilde{S}_r^{z}}\big) \tilde{S}_q^{z}
\Big\rangle$~\cite{denNijsRom1989PRB}. Notably, $M_L$ and $\mathcal{O}_H$ are
measured in the original and rotated bases, respectively.

We start by examining a vertical cut at $D = -0.7$. Both $M_L$ and
$\mathcal{O}_H$ drop sharply to nearly zero at $\vartheta_c/\pi = -0.4909(2)$
(Supplementary Fig.~\ref{fig:SMd070}), indicative of a first-order transition. We next examine
the vertical cut at $D=-1.0$, where the dimer order parameter exhibits a smooth
evolution with system size, indicating a continuous phase transition. 
To characterize the transition, we compute the Binder cumulant 
$U_L=1-\langle\hat M_L^4\rangle/(3\langle\hat M_L^2\rangle^2)$, 
which obeys the finite-size scaling ansatz
\begin{align}\label{eq:binder}
U_L = f\bigl(L^{1/\nu}(\vartheta-\vartheta_c),\, L^{-\omega}\bigr),
\end{align}
where $\nu$ is the correlation-length exponent and $\omega$ accounts for scaling
corrections~\cite{Shao2016Sci}. As shown in Fig.~\ref{fig:binder}a, the
intersections of $U_L$ for different system sizes are close but do not coincide
exactly. We therefore perform an extrapolation using the crossing points of pairs
$(L,2L)$, plotted in Fig.~\ref{fig:binder}c, which yields the critical point
$\vartheta_c/\pi = -0.475898(2)$ in the thermodynamic limit. Furthermore, from the
scaling relation~\cite{Liu2019NC}
\begin{align}\label{eq:nu}
\frac{1}{\nu} = \frac{1}{\ln 2} \ln\left(\frac{\partial U_{2L}/\partial \vartheta}{\partial U_{L}/\partial \vartheta}\right)_{\vartheta_c},
\end{align}
we obtain $1/\nu \approx 1.25$, i.e., $\nu \approx 0.80$
[see Fig.~\ref{fig:binder}b].

Independently, we extract the scaling dimension $d = \beta/\nu$ via boundary
conformal field theory, where the local dimerization decays as
$m_L(l) \propto \left[\frac{L}{\pi}\sin\left(\frac{\pi l}{L}\right)\right]^{-d}$
~\cite{Chepiga2016PRBa}. Fitting the log-linear behavior of $m_L(l)$ against the
conformal distance $r_L = \sin(\pi l/L)$ for $l\in[L/4,L/2]$ yields the scaling
dimensions $d \approx 0.297, 0.294$, and $0.293$ for system sizes $L = 128, 192,$
and $256$, respectively, at $\vartheta_c$, see
Figs.~\ref{fig:dimension}a--c. The nearly identical values across different
$L$ indicate weak finite-size effects. Extrapolating these values using a
quadratic fit in $1/L$ gives $d_{\infty} \approx 0.291(1)$, which serves as our
best estimate of the scaling dimension in the thermodynamic limit at this
parameter point.

The two cuts at $D=-0.7$ and $D=-1.0$ therefore place the same boundary on
opposite sides of a changeover (Fig.~\ref{fig:phase}): it is discontinuous close to the transition point, 
where the two order parameters exchange their weight abruptly, and continuous at $D=-1.0$,
where a finite correlation-length critical exponent and a well-defined scaling dimension can be
extracted. The changeover itself is located at
$(D_c,\vartheta_c/\pi) \approx (-0.73,-0.4893)$, marked by the black dot in the inset of
Fig.~\ref{fig:phase}, where the first-order segment terminates and the continuous
segment begins. It remains to identify the critical theory that governs the continuous
segment.

\begin{figure}[htbp]
\centering
\includegraphics[width=0.95\columnwidth, clip]{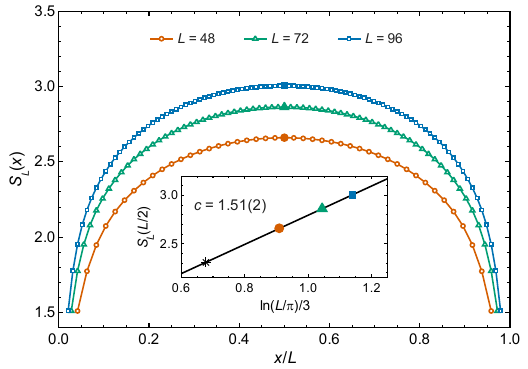}
\caption{\textbf{Entanglement entropy and central charge at the critical
point.} Von Neumann entropy $\mathcal{S}_{L}(x)$ as a function
of the partition size $x/L$ at the quantum critical point $\vartheta_{c}/\pi$ in
the Kitaev--$\Gamma$ spin chain with SIA $D=-1.0$ for chain lengths $L=48$ (red
circle), $72$ (green triangle), and $96$ (blue square). Inset, estimate of the
central charge $c=1.51(2)$.}
  \label{fig:cc}
\end{figure}

\artsub{Emergent $\mathrm{SU}(2)_2$ Wess--Zumino--Witten criticality and bosonization}

The entanglement-entropy scaling at this critical point yields a central charge
$c = 1.51(2)$ (Fig.~\ref{fig:cc}). With the numerical values of $\nu$, $d = \beta/\nu$, and $c$ at hand, we
now compare them against the theoretical predictions for the
$\mathrm{SU}(2)_2$ WZW universality class. Affleck and Haldane conjectured that,
in the absence of explicit dimerization, the continuous transition from the
Haldane phase to a dimerized phase in spin-1 chains should be described by the
$\mathrm{SU}(2)_{k=2S}$ WZW model~\cite{Affleck1987PRB}. For $S=1$ (i.e., $k=2$),
this scenario predicts $c= 3k/(2+k) = 3/2$, $\nu = (2+k)/(2k) = 1$, and
$d=\beta/\nu = 3/[2(2+k)] = 3/8$. Our extracted central charge $c\approx 1.5$ is
in precise agreement with this prediction, lending strong support to the WZW
identification.

However, the exponents we extract at $D=-1$, namely $\nu\approx 0.80$ and
$d \approx 0.291$, deviate from the WZW fixed-point predictions. Rather than
contradicting the scenario, this deviation signals a marginal critical flow along
the phase boundary: the presence of a marginal operator, whose strength is tuned
by the SIA $D$, renormalizes the effective exponents away from their universal
fixed-point values. To test this hypothesis, we track the extracted scaling
dimension $d$ along the entire critical line. As shown in
Fig.~\ref{fig:dimension}d, $d(D)$ increases monotonically with $|D|$ and,
notably, crosses the exact $\mathrm{SU}(2)_2$ WZW value of $3/8$ at
$D \approx -2.5$. Therefore, while $D = -1$ exhibits a clear WZW central charge,
its critical exponents are renormalized by the marginal flow; the pristine
$\mathrm{SU}(2)_2$ WZW fixed point is only asymptotically reached at
$D \approx -2.5$.

\begin{figure}[htbp]
\centering
\includegraphics[width=0.95\columnwidth, clip]{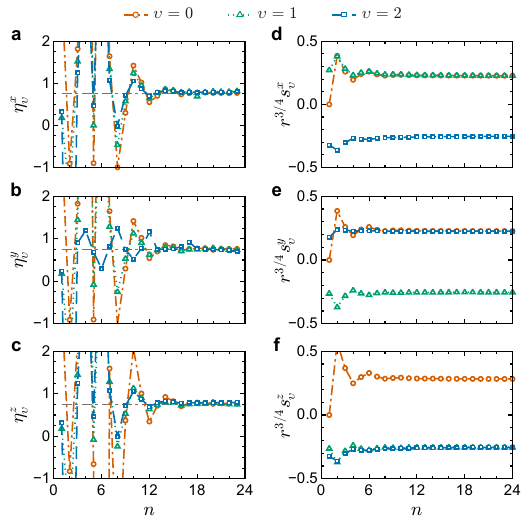}
\caption{\textbf{Universal critical exponent $\eta$ from the asymptotic spin
correlators.} \textsf{\textbf{a}}--\textsf{\textbf{c}}, Estimates of
$\eta_{\upsilon}^{\gamma}$ from Eq.~\eqref{eq:eta} for the Kitaev--$\Gamma$ spin
chain with SIA $D=-1.0$ at chain length $L=144$ under PBC, shown for
\textsf{\textbf{a}}, $\gamma=x$; \textsf{\textbf{b}}, $\gamma=y$; and
\textsf{\textbf{c}}, $\gamma=z$. In each panel, the horizontal dashed line marks
the theoretical value $\eta=3/4$, and different curves denote components
$\upsilon=0$ (red circles), $\upsilon=1$ (green triangles), and $\upsilon=2$
(blue squares). \textsf{\textbf{d}}--\textsf{\textbf{f}}, The corresponding
$r^{3/4} s_{\upsilon}^{\gamma}$ for \textsf{\textbf{d}}, $\gamma=x$;
\textsf{\textbf{e}}, $\gamma=y$; and \textsf{\textbf{f}}, $\gamma=z$.}
  \label{fig:eta}
\end{figure}

The two observations are not in conflict, because the central charge and the
scaling dimensions respond differently to the same deformation. The central
charge counts the gapless degrees of freedom and is relatively insensitive to 
marginal perturbations, whereas $\nu$ and $d$ measure the distance to the fixed point along the
resulting trajectory. A matching $c$ accompanied by drifting exponents should
accordingly be read as a flow towards the universality class rather than as
evidence against it---a distinction that matters whenever a fixed point is
approached only asymptotically, and one that exponent matching alone cannot
resolve.

Crucially, a bosonization analysis (detailed in Methods) predicts the asymptotic
spin correlation
\begin{align}\label{eq:corr}
\langle \tilde{S}_1^\gamma \tilde{S}_{1+r}^\gamma \rangle = -E_{D,\upsilon}^{\gamma} \frac{1}{r^2} + (-1)^r E_{C,\upsilon}^{\gamma} \frac{\ln^{\rho}(r/r_0)} {r^{\eta}},
\end{align}
in which $r$ denotes distance, $r_0$ is a nonuniversal length scale, and
$E_{D,\upsilon}^\gamma, E_{C,\upsilon}^\gamma$ are bond-dependent coefficients for
the uniform and staggered parts, respectively, where $\upsilon \equiv r\mod 3$
($0\leq \upsilon \leq 2$). For the $\mathrm{SU}(2)_2$ WZW universality class, one
expects $\eta = 3/4$. Fitting our numerical data at $D = -1.0$ directly yields
$\eta \approx 0.77(3)$ (Fig.~\ref{fig:eta}), in excellent quantitative
agreement with this prediction.
We note that the scaling relation $\beta/\nu = \eta/2$ would imply $\beta/\nu = 3/8$
at the WZW fixed point; our independently extracted $\beta/\nu$ from finite-size
scaling shows a slight deviation, which we attribute to the marginal flow along
the critical line. Nevertheless, the direct estimate of $\eta$ from the
correlation decay is robust and provides compelling, self-consistent evidence that
the continuous dimer--Haldane transition indeed belongs to the
$\mathrm{SU}(2)_2$ WZW universality class.


\artsec{Discussion}

Our results show that single-ion anisotropy controls both the character of the
gapped regimes and the nature of the dimer--Haldane transition in the spin-1
Kitaev--$\Gamma$ chain. On the positive side, we identify a smooth crossover from
the Kitaev phase to the large-$D$ phase, evidenced by the quadratic-to-linear gap
evolution, coexisting spin-nematic and string order parameters, and the
disappearance of the double-peak specific heat. On the negative side, increasing
$|D|$ changes the dimer--Haldane transition from first order to continuous. The
central finding is a continuous transition with numerical signatures consistent
with $\mathrm{SU}(2)_2$ WZW criticality despite the absence of microscopic
continuous spin-rotation symmetry: the agreement of the central charge and
spin-correlation exponent with $c=3/2$ and $\eta=3/4$ provides complementary
evidence for this identification. The change from first-order to continuous
behavior further demonstrates that anisotropy can tune access to this emergent
critical regime.

This finding connects two previously established settings: emergent SU(2)
symmetry in the spin-$1/2$ Kitaev--$\Gamma$ chain~\cite{Yang2020PRL} and
$\mathrm{SU}(2)_2$ dimerization transitions in spin-rotation-invariant spin-1
chains~\cite{Chepiga2016PRBa,Chepiga2016PRBc}. The present model brings these
features together at a transition involving the topological Haldane phase. In the
rotated frame, the symmetry-based bosonization analysis supports an
SU(2)-invariant low-energy description: once the three-site unit cell can no
longer be resolved at long distances, the exchanges combine into an
$\mathrm{SU}(2)$-invariant form and the anisotropy reduces to a term quadratic in
the smeared spin, so that the low-energy theory acquires a symmetry that the
microscopic Hamiltonian does not possess (Supplementary
Note~\ref{SM:U6Boson}). At the same time, sublattice-dependent correlation
amplitudes retain information about the microscopic discrete symmetries. This
distinction between universal scaling and microscopic amplitudes helps explain
how high-symmetry critical behavior can emerge from bond-directional
interactions.

The finite-size estimates of the dimerization scaling dimension and
correlation-length exponent require further consideration. At $D=-1$, the
extracted values $d\approx0.291$ and $\nu\approx0.80$ differ appreciably from the
WZW predictions $d=3/8$ and $\nu=1$. Marginal perturbations allowed by the
low-energy theory offer a possible explanation through logarithmic corrections to
scaling. The variation of the effective dimerization exponent along the phase
boundary motivates this interpretation, but does not by itself establish the
corresponding renormalization-group flow. Explicit treatment of the correction
exponent $\omega$ and of the logarithmic corrections, together with
larger-system calculations or low-energy spectral analysis, would provide a more
stringent test.

The positive-$D$ crossover offers a complementary perspective: pronounced changes
in excitation and thermal signatures can occur within an adiabatically connected
gapped phase. In particular, the disappearance of the low-temperature
specific-heat peak characterizes the evolution towards the large-$D$ limit
without defining a thermodynamic phase boundary. Together, the crossover and the
tunable dimer--Haldane transition make this chain a useful setting for
distinguishing changes in observable signatures from changes in phase and
universality class. Dynamical spin correlations and the finite-size excitation
spectrum provide natural next tests of the proposed emergent WZW description,
and it will be interesting to examine whether the same mechanism---an emergent
symmetry selecting a WZW fixed point in the absence of microscopic
$\mathrm{SU}(2)$ symmetry---operates in other bond-directional chains and
higher-spin generalizations. More broadly, our findings identify
bond-directional exchange as a route to emergent conformal criticality, 
opening an avenue to WZW physics in quantum magnets unconstrained
by continuous symmetry.


\artsec{Methods}

\artsub{Many-body computations}

Ground-state calculations were performed with the variational DMRG
algorithm~\cite{White1992,Schollwock2005}. We routinely keep 2000 states (up to
4000 occasionally), with truncation errors below $\sim10^{-7}$. Excitation gaps
were obtained from the energy difference between the ground state and the lowest
excited state, and only data that are stable with
respect to the number of kept states are reported. Order parameters were evaluated
in the central region of the chain to avoid edge effects. PBC are used for gaps
and for the correlation functions employed in the bosonization analysis, and OBC
for the order parameters and Binder cumulant.

Thermodynamic quantities were computed with the transfer-matrix formulation of
the DMRG at finite temperature~\cite{BurXiangGeh1996,WangXiang1997}. The specific
heat follows from the internal energy as $C_v=\partial U/\partial T$, and the loci
of its extrema were determined from the condition $\partial C_v/\partial T=0$ on a
grid of temperature and anisotropy, with spacings of $0.01$ in $D$ and $0.002$ in
$T$ in the region where the two low-temperature extrema approach each other. This
grid fixes the merging point $D=0.98(1)$ quoted in the main text. The same
procedure was applied at $\vartheta/\pi=-0.45$, $-0.40$, and $-0.35$
(Supplementary Fig.~\ref{fig:SMcv}).

\artsub{Finite-size scaling analysis}

The Binder cumulant of the dimer order parameter was computed for chain lengths
$L=32$, $48$, $64$, $96$, $128$, and $192$ at $D=-1.0$. The critical point was
obtained from the crossing points $\vartheta_c(L,2L)$ of successive pairs of
sizes, extrapolated in $1/L$, which gives $\vartheta_c/\pi=-0.475898(2)$; the
inverse correlation-length exponent follows from Eq.~\eqref{eq:nu}, evaluated from
the slopes of $U_L$ at the crossing, giving $1/\nu\approx1.25$ and
$\nu\approx0.80$.

The scaling dimension was extracted from the decay of the local dimerization at
criticality as given in the main text, by fitting $\ln m_L(l)$
against $\ln\sin(\pi l/L)$ over $l\in[L/4,L/2]$. This window excludes the
edge region, where the open ends perturb the local dimerization. The values
$d\approx0.297$, $0.294$, and $0.293$ for $L=128$, $192$,
and $256$ were extrapolated quadratically in $1/L$, giving
$d_\infty\approx0.291(1)$. The same procedure, applied along the critical line for
$D\in[-3,-1]$, produces the function $d(D)$ of Fig.~\ref{fig:dimension}d.

The central charge was obtained from the von Neumann entanglement entropy. At
criticality, the entropy of a subsystem of length $x$ in a periodic chain of size
$L$ obeys
\begin{align}\label{eq:entropy}
\mathcal{S}_L(x)=\frac{c}{3}\ln\left[\frac{L}{\pi}\sin\left(\frac{\pi x}{L}\right)\right] + c',
\end{align}
with $c$ the central charge and $c'$ a nonuniversal
constant~\cite{Calabrese2004JSM}. We compute
$\mathcal{S}_L(x)$ for $L=48,72,96$ [Fig.~\ref{fig:cc}] and extract $c$ from fits
to the $x$-dependence, excluding boundary points. The extracted values are
$c_L=1.509,1.503,1.497$ for $L=48,72,96$, respectively, which are reasonably close
to $3/2$. A separate fit using the half-chain entropy
$\mathcal{S}_L(L/2)=\frac{c}{3}\ln(L/\pi)+c'$ yields $c=1.51(2)$ in the
thermodynamic limit (inset). Both analyses consistently yield $c=3/2$.

\artsub{Bosonization of the low-energy theory}

At low energies, it is enough to keep only the relevant and marginal operators in
the $\mathrm{SU}(2)_2$ WZW model, which are consistent with the symmetries of the
Kitaev--$\Gamma$ chain with a uniaxial SIA in the $U_6$ frame (for detailed
discussions of the symmetries, see Supplementary Note~\ref{SM:U6Boson}). The low energy
Hamiltonian contains a relevant and a marginal term, written as
\begin{align}\label{eq:low}
\mathcal{H}_{\text{Low}} = \mathcal{H}_{\text{WZW}} +\int dx \big[
\lambda_1(\text{tr}\,g)^2+\lambda_2 \vec{J}_\mathcal{L}\cdot \vec{J}_\mathcal{R}
\big],
\end{align}
where $g$ is the primary field, and $\vec{J}_\mathcal{L}$ and $\vec{J}_\mathcal{R}$
are the left and right WZW current operators. Since the low energy dimerization
operator is given by $\text{tr}(g)$~\cite{Affleck1987PRB}, the system develops a
dimerization order when $\lambda_1<0$, corresponding to the dimer phase, and does
not have a dimerization order when $\lambda_1>0$, corresponding to the Haldane
phase. Right on the critical line, $\lambda_1$ vanishes. The marginal coupling
$\lambda_2$ survives on that line, and it is the term responsible for the flow of
the critical exponents discussed in the main text.

Along the critical line, although there is an emergent SU(2) symmetry at low
energies, the system still has reminiscence of the exact discrete symmetries even
in the long distance limit~\cite{Yang2020PRL}. The bosonization formulas consistent
with the exact symmetries of the system can be derived as~\cite{Yang2020PRL}
\begin{align}\label{eq:bos}
\tilde{S}_{\upsilon+3n}^{ \gamma}=D_{\upsilon}^\gamma(J_\mathcal{L}^\gamma+J_\mathcal{R}^\gamma)+(-1)^{\upsilon+3n} C_{\upsilon}^{\gamma} \text{tr} (g\sigma^\gamma),
\end{align}
where $0\leq \upsilon \leq 2$, $\gamma\in\{x,y,z\}$, and the
$C_{\upsilon}^\gamma,D_{\upsilon}^{\gamma}$ coefficients satisfy the symmetry
constraints $\Lambda_0^x=\Lambda_1^z=\Lambda_2^y (=\Lambda_1)$ and
$\Lambda_0^y=\Lambda_1^x=\Lambda_2^z=\Lambda_0^z=\Lambda_1^y=\Lambda_2^x (=\Lambda_2)$
where $\Lambda=C, D$.

Upon proper normalizations, the equal-time correlation functions between WZW
currents and primary fields are given by
$\left<J_\mathcal{L}^\gamma(0)J_\mathcal{L}^\gamma(r)\right> = -\frac{1}{2r^2}$,
$\left<J_\mathcal{R}^\gamma(0)J_\mathcal{R}^\gamma(r)\right> = -\frac{1}{2r^2}$, and
$\left<N^\gamma(0)N^\gamma(r)\right> = \frac{[\ln (r/r_0)]^\rho}{r^{3/4}}$,
where $\rho = 1/2$ at the $\mathrm{SU}(2)_2$ WZW fixed point.
Hence, the equal-time spin correlation functions in the $r\gg 1$ limit can be
derived as
\begin{align}\label{eq:CorrYW}
\langle \tilde{S}_1^\gamma \tilde{S}_{1+\upsilon+3n}^\gamma \rangle = -E_{D,\upsilon}^{\gamma}\frac{1}{r^2} + (-1)^{\upsilon+3n} E_{C,\upsilon}^{\gamma} \frac{\ln^{\frac{1}{2}}(r/r_0)} {r^{3/4}},
\end{align}
in which $E_{\Lambda,\upsilon}^{\gamma}$ ($\Lambda=C,D$) can be determined as
\begin{align}
\begin{array}{|c|c|c|c|}
\hline
E_{\Lambda,\upsilon}^\gamma & \gamma=x & \gamma=y & \gamma=z\\
\hline
\upsilon=0 & (\Lambda_2)^2 & (\Lambda_2)^2 & (\Lambda_1)^2\\
\hline
\upsilon=1 & (\Lambda_2)^2 & \Lambda_1\Lambda_2 & \Lambda_1\Lambda_2 \\
\hline
\upsilon=2 & \Lambda_1\Lambda_2 &  (\Lambda_2)^2 & \Lambda_1\Lambda_2\\
\hline
\end{array}
\label{eq:etable}
\end{align}
On finite-size systems with PBC, $r$ in Eq.~\eqref{eq:CorrYW} has to be replaced with
$r_L=\frac{L}{\pi}\sin(\frac{\pi r}{L})$, where $L$ is the number of sites in the
system.

To extract the staggered spin correlations, the DMRG data are binned modulo 3 with
fixed $\upsilon\in\{0,1,2\}$. The correlation function
$f_{\upsilon}^\gamma(n)=\langle \tilde{S}_1^{\gamma} \tilde{S}_{1+\upsilon+3n}^{\gamma}\rangle$
is decomposed into a uniform background and a staggered component as
$f_{\upsilon}^\gamma(n) = u_{\upsilon}^\gamma(n) +(-1)^{\upsilon+3n} s_{\upsilon}^\gamma(n)$,
where $u_{\upsilon}^\gamma(n)=-E_{D,\upsilon}^\gamma r_L^{-2}$ and
$s_{\upsilon}^\gamma(n)=E_{C,\upsilon}^\gamma \ln^{1/2}(r_L/r_0) r_L^{-3/4}$, with
$r_L=\frac{L}{\pi}\sin(\frac{\pi(\upsilon+3n)}{L})$. The critical exponent $\eta$ is
obtained via
\begin{align}\label{eq:eta}
\eta_{\upsilon}^{\gamma} = 2\frac{[(\ln s_{\upsilon}^{\gamma})'']^2}{(\ln s_{\upsilon}^{\gamma})'''} - (\ln s_{\upsilon}^{\gamma})',
\end{align}
where primes denote derivatives with respect to $\ln r = \ln(\upsilon+3n)$. As
shown in Figs.~\ref{fig:eta}a, b, and c for $\upsilon=0,1,2$, respectively,
the long-distance extrapolation yields $\eta=0.77(3)$, consistent with $3/4$. To
further verify this scaling, we plot $r^{3/4} s_{\upsilon}^{\gamma}$ in
Figs.~\ref{fig:eta}d, e, and f for $\upsilon=0,1,2$, respectively; the nearly
site-independent collapse gives $C_1 \approx 0.5319$ and $C_2 \approx 0.4766$.


\artsec{Data availability}
The data that support the findings of this article are openly available in the
Zenodo repository at \href{https://doi.org/10.5281/zenodo.24307086}{https://doi.org/10.5281/zenodo.24307086}
(ref.~\cite{DataAndCode}).

\artsec{Code availability}
All numerical codes in this paper are available upon request to the
authors.

\artsec{Acknowledgements}
We thank S. Eggert for helpful discussions. The computations are partially
supported by the High Performance Computing Platform of Nanjing University of
Aeronautics and Astronautics (NUAA).

\artsec{Author contributions}
Q.L. and X.W. conceived the project. Q.L. performed the density-matrix
renormalization group calculations and the data analysis. S.H. guided the
finite-temperature calculations. W.Y.
carried out the bosonization analysis. Y.L., J.L., and J.Z. took part in
scientific discussions throughout the project. Q.L. and W.Y. wrote the paper
and all authors helped revise the manuscript.

\artsec{Funding}
This work is supported by the National Program on Key Research Project (Grant
No.\ MOST2022YFA1402700), the National Natural Science Foundation of China
(Grants Nos.\ 12674185, 12674178, 12574163, 12274187, 12305039, and
12474476), and the Beijing National Laboratory for Condensed Matter Physics
(Grant No.\ 2025BNLCMPKF022).

\artsec{Competing interests}
The authors declare no competing interests.

\artsec{Additional information}
\textbf{Supplementary information} is available in the online version of the paper.\\
\textbf{Correspondence} and requests for materials should be addressed to Q.L.,
W.Y., or X.W.



\clearpage
\onecolumngrid

\newpage
\setcounter{page}{1}

\setcounter{equation}{0}
\setcounter{figure}{0}
\setcounter{table}{0}
\setcounter{smnote}{0}

\renewcommand{\theequation}{S\arabic{equation}}
\renewcommand{\thefigure}{\arabic{figure}}
\renewcommand{\thetable}{S\arabic{table}}
\renewcommand{\theHequation}{suppEq\arabic{equation}}
\renewcommand{\theHfigure}{suppFig\arabic{figure}}
\renewcommand{\theHtable}{suppTab\arabic{table}}
\renewcommand{\bibnumfmt}[1]{[#1]}
\renewcommand{\citenumfont}[1]{#1}

\begin{center}
{\large{\bf Supplemental Information for\\
 ``Emergent $\mathrm{SU}(2)_2$ Wess--Zumino--Witten criticality\\ in an anisotropic spin-1 Kitaev--$\Gamma$ chain''}}
\end{center}
\begin{center}
Qiang Luo,$^{1,2}$ Shijie Hu,$^{3}$ Wang Yang,$^{4}$ Yuhai Liu,$^{5}$ Jinbin Li,$^{1,2}$ Jize Zhao,$^{6}$ and Xiaoqun Wang$^{7}$\\
\quad\\
$^1$\textit{College of Physics, Nanjing University of Aeronautics and Astronautics, Nanjing 211106, China}\\
$^2$\textit{Key Laboratory of Aerospace Information Perception and Physics (NUAA), Nanjing 211106, China}\\
$^3$\textit{Beijing Computational Science Research Center, Beijing 100084, China}\\
$^4$\textit{School of Physics, Nankai University, Tianjin 300071, China}\\
$^5$\textit{School of Physical Science and Technology, Beijing University of Posts and Telecommunications, Beijing 100876, China}\\
$^6$\textit{School of Physical Science and Technology \& Lanzhou Center for\\ Theoretical Physics, Lanzhou University, Lanzhou 730000, China}\\
$^7$\textit{School of Physics, Zhejiang University, Hangzhou 310058, China}\\
(Dated: November 18th, 2026)
\quad\\
\end{center}


In this Supplemental Information, we present the six-sublattice rotation and the
bosonization analysis of the effective Hamiltonian for the dimer--Haldane
transition (Supplementary Note~\ref{SM:U6Boson}), the evolution of the double-peak
structure in specific heat across the Kitaev--large-$D$ crossover (Supplementary
Note~\ref{SM:SpecHeat}), the finite-size scaling of the dimer order parameter at
the dimer--Kitaev transition (Supplementary Note~\ref{SM:FSS}), and the
order-parameter jumps at $D=-0.7$ that identify the first-order character of the
dimer--Haldane transition (Supplementary Note~\ref{SM:D070}).

\smnote{Six-sublattice rotation and bosonization analysis}{SM:U6Boson}

The six-sublattice rotation $U_6$ is defined as
\begin{eqnarray}
\text{Sublattice $1$}: & (x,y,z) & \rightarrow (-\tilde{x},-\tilde{y},\tilde{z})\nonumber\\
\text{Sublattice $2$}: & (x,y,z) & \rightarrow (\tilde{x},\tilde{z},-\tilde{y})\nonumber\\
\text{Sublattice $3$}: & (x,y,z) & \rightarrow (-\tilde{y},-\tilde{z},\tilde{x})\nonumber\\
\text{Sublattice $4$}: & (x,y,z) & \rightarrow (\tilde{y},\tilde{x},-\tilde{z})\nonumber\\
\text{Sublattice $5$}: & (x,y,z) & \rightarrow (-\tilde{z},-\tilde{x},\tilde{y})\nonumber\\
\text{Sublattice $6$}: & (x,y,z) & \rightarrow (\tilde{z},\tilde{y},-\tilde{x}),
\label{eq:6rotation}
\end{eqnarray}
in which ``Sublattice $i$'' ($1\leq i \leq 6$) represents all the sites $i+6n$
($n\in\mathbb{Z}$) in the chain, $\gamma$ and $\tilde{\gamma}$ ($\gamma=x,y,z$)
are $S^\gamma$ before $U_6$ rotation and $\tilde{S}^{\gamma}$ after $U_6$ rotation
for short. It can be verified that the Hamiltonian $\mathcal{H}^\prime$ after
$U_6$ has a three-site periodicity, having the following form in a unit cell,
\begin{eqnarray}
\mathcal{H}^\prime_{1+3n,2+3n}&=&-K\tilde{S}_{1+3n}^{x}\tilde{S}_{2+3n}^{x}
+\Gamma(\tilde{S}_{1+3n}^{y}\tilde{S}_{2+3n}^{y}
+\tilde{S}_{1+3n}^{z}\tilde{S}_{2+3n}^{z})
+D(\tilde{S}_{1+3n}^{z})^2\nonumber\\
\mathcal{H}^\prime_{2+3n,3+3n}&=&-K\tilde{S}_{2+3n}^{z}\tilde{S}_{3+3n}^{z}
+\Gamma(\tilde{S}_{2+3n}^{x}\tilde{S}_{3+3n}^{x}
+\tilde{S}_{2+3n}^{y}\tilde{S}_{3+3n}^{y})
+D(\tilde{S}_{2+3n}^{y})^2\nonumber\\
\mathcal{H}^\prime_{3+3n,4+3n}&=&-K\tilde{S}_{3+3n}^{y}\tilde{S}_{4+3n}^{y}
+\Gamma(\tilde{S}_{3+3n}^{z}\tilde{S}_{4+3n}^{z}
+\tilde{S}_{3+3n}^{x}\tilde{S}_{4+3n}^{x})
+D(\tilde{S}_{3+3n}^{x})^2.
\label{eq:Hprime}
\end{eqnarray}

The Hamiltonian $\mathcal{H}^\prime$ is invariant under the following symmetry
operations,
\begin{eqnarray}
1.&\mathcal{T} &:  (\tilde{S}_i^{x},\tilde{S}_i^{y},\tilde{S}_i^{z})
\rightarrow (-\tilde{S}_{i}^{x},-\tilde{S}_{i}^{y},-\tilde{S}_{i}^{z})\nonumber\\
2.& R(\frac{1}{\sqrt{3}}(1,1,1),\frac{2\pi}{3})T_{a}&:  (\tilde{S}_i^{x},\tilde{S}_i^{y},\tilde{S}_i^{z})
\rightarrow (\tilde{S}_{i+1}^{z},\tilde{S}_{i+1}^{x},\tilde{S}_{i+1}^{y})\nonumber\\
3.&R(\frac{1}{\sqrt{2}}(1,0,-1)) I&: (\tilde{S}_i^{x},\tilde{S}_i^{y},\tilde{S}_i^{z})
\rightarrow (-\tilde{S}_{4-i}^{z},-\tilde{S}_{4-i}^{y},-\tilde{S}_{4-i}^{x})\nonumber\\
4.&R(\hat{x},\pi) &:  (\tilde{S}_i^{x},\tilde{S}_i^{y},\tilde{S}_i^{z})
\rightarrow (\tilde{S}_i^{x},-\tilde{S}_i^{y},-\tilde{S}_i^{z})\nonumber\\
5.&R(\hat{y},\pi) &:  (\tilde{S}_i^{x},\tilde{S}_i^{y},\tilde{S}_i^{z})
\rightarrow (-\tilde{S}_i^{x},\tilde{S}_i^{y},-\tilde{S}_i^{z})\nonumber\\
6.&R(\hat{z},\pi) &:  (\tilde{S}_i^{x},\tilde{S}_i^{y},\tilde{S}_i^{z})
\rightarrow (-\tilde{S}_i^{x},-\tilde{S}_i^{y},\tilde{S}_i^{z}),
\label{eq:symmetries_Ez}
\end{eqnarray}
in which $\mathcal{T}$ is time reversal, $T_a$ is the spatial translation operator
by one lattice site, $I$ is spatial inversion with inversion center at site $2$,
and $R(\hat{n},\varphi)$ is the global spin rotation around $\hat{n}$-axis by an
angle $\varphi$. The symmetry group $G$ is generated by the operations in
Eq.~(\ref{eq:symmetries_Ez}), which has been shown to exhibit a nonsymmorphic
$O_h$ group structure in Ref.~\cite{SMYang2020PRL}, where $O_h$ is the full
octahedral group. It can be shown that Eq.~\eqref{eq:bos} of Methods is the most
general formula which is consistent with the symmetries in
Eq.~(\ref{eq:symmetries_Ez}).

\begin{figure}[tbp]
\centering
\includegraphics[width=0.95\textwidth, clip]{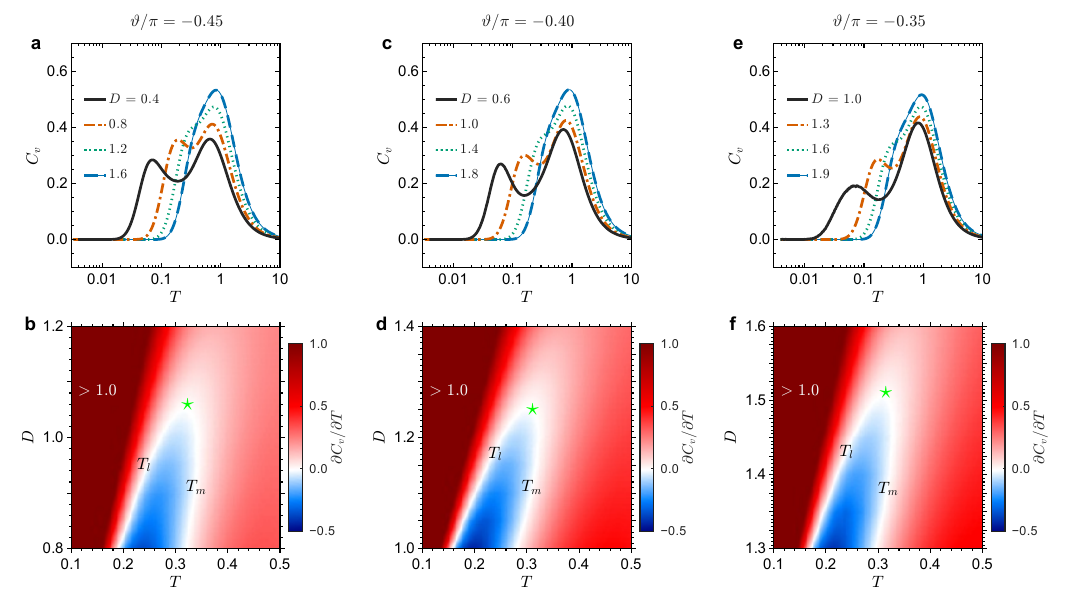}
\renewcommand{\figurename}{\textbf{Supplementary Fig.}}
\caption{\textbf{Double-peak specific heat across the Kitaev-to-large-$D$
crossover.} \textbf{a},\textbf{c},\textbf{e},
Specific heat $C_v$ versus temperature $T$ for the Kitaev--$\Gamma$ chain with
SIA, shown for $\vartheta/\pi=-0.45$ (with $D=0.4,0.8,1.2,1.6$), $-0.40$ (with
$D=0.6,1.0,1.4,1.8$), and $-0.35$ (with $D=1.0,1.3,1.6,1.9$).
\textbf{b},\textbf{d},\textbf{f}, Corresponding color maps of $\partial C_v/\partial T$ in
the $(T,D)$ plane; $T_l$ and $T_m$ mark the loci of $\partial C_v/\partial T = 0$,
with the stars marking where the low-$T$ peaks disappear.}
\label{fig:SMcv}
\end{figure}

We provide an intuitive understanding as to why there is an emergent SU(2)
symmetry along the critical line between the Haldane and dimer phases. In the
six-sublattice rotated frame, if the system is in a critical state, then at long
distances, the spin operators on different sites get smeared so that adjacent
sites can no longer be clearly distinguished. Neglecting the differences in the
site indices in Eq.~(\ref{eq:Hprime}) and summing over all interactions within a
unit cell, we obtain the following smeared Hamiltonian,
\begin{eqnarray}
\mathcal{H}^\prime_{\text{smear}}&\sim& \int dx\big[ (-K+2\Gamma) \vec{S}^\prime(x)\cdot \vec{S}^\prime(x+1)+D\vec{S}^\prime(x)\cdot \vec{S}^\prime(x)\big],
\label{eq:Hp_continuum}
\end{eqnarray}
in which $\vec{S}^\prime(x)$ is the smeared spin operator at position $x$ in the
continuum limit. Clearly, $\mathcal{H}^\prime_{\text{smear}}$ in
Eq.~(\ref{eq:Hp_continuum}) is SU(2) invariant, which explains intuitively why an
SU(2) symmetry emerges at low energies. Therefore, it is reasonable to expect that
the low energy critical theory is in the same universality class as the one for
the continuous phase transitions between Haldane and dimer phases in SU(2)
invariant spin-1 chains, which has been established to be the
$\mathrm{SU}(2)_2$ Wess--Zumino--Witten (WZW)
model~\cite{SMChepiga2016PRBa,SMAffleck1987PRB}.

We briefly comment on how the bosonization formulas in Eq.~\eqref{eq:bos} of
Methods are obtained in the $U_6$ frame. Following Ref.~\cite{SMAffleck1987PRB},
the low energy physics of spin-$S$ chains can be mimicked by a generalized Hubbard
model of electrons in the weak coupling limit, where the number of electrons is
$2S$ per site and the interaction is the on-site Hund's interaction. The reduction
from spin model to interacting fermion model is based on the observation that
there is no phase transition in the fermion model from weak-$U$ to strong-$U$
limit and the fact that spin models can be recovered from fermion model in the
strong coupling limit~\cite{SMAffleck1987PRB}. For this reason, we consider
bosonization formulas in the interacting fermion model.

Linearizing the fermion spectrum around the Fermi points, the low energy degrees
of freedom in the fermion model are confined to the neighborhoods of left and
right Fermi points, having linear dispersions and emergent Lorentz symmetry. In
the low energy limit, the free fermion model can be bosonized into a combination
of U(1) charge boson, an SU(2) spin boson, and a SU(2S) color boson. Interactions
in general gap out the charge and color boson sectors at low
energies~\cite{SMAffleck1987PRB}. It can be shown that in the fermion model, the
intra-Fermi-point contributions to spin operators are the WZW current operators
in the spin boson sector, giving rise to the $J_\mathcal{L}^\gamma,
J_\mathcal{R}^\gamma$ terms in Eq.~\eqref{eq:bos}. For the inter-Fermi-point
contributions, $\text{tr} (g\sigma^\gamma)$ in Eq.~\eqref{eq:bos} is the operator
which has the correct symmetry transformation property and the smallest scaling
dimension. Therefore, by only retaining the leading operators, the bosonization
formulas for spin operators can be captured by Eq.~\eqref{eq:bos}. The coefficients
$D_{\upsilon}^\gamma,C_{\upsilon}^\gamma$ arise from the wave function
renormalization effects induced by SU(2) breaking terms in the Hamiltonian in the
$U_6$ frame, as discussed for the spin-$1/2$ case in Ref.~\cite{SMYang2020PRL}.

\smnote{Rise and fall of the double-peak structure in specific heat}{SM:SpecHeat}

As a complement to the discussion of $\vartheta/\pi = -0.50$ in the main text, we
present here the evolution of the double-peak structure in the specific heat at
selected exchange-coupling angle $\vartheta$ for $\vartheta/\pi = -0.45$ [see
Supplementary Fig.~\ref{fig:SMcv}a, b], $-0.40$ [see Supplementary Fig.~\ref{fig:SMcv}c, d], and
$-0.35$ [see Supplementary Fig.~\ref{fig:SMcv}e, f]. For small $D$, the specific heat
exhibits a clear double-peak structure. As $D$ increases, the low-temperature peak
gradually smears out and eventually disappears. This universal behavior indicates
a significant difference in the low-energy excitations between the Kitaev phase
in the small-$D$ regime and the large-$D$ phase.

\smnote{Finite-size scaling of dimer order parameter}{SM:FSS}

The dimer--Kitaev transition at the Kitaev-only coupling
($\vartheta/\pi = -1/2$) belongs to the Ising universality class and remains so
upon inclusion of the $\Gamma$ interaction. To confirm this, we perform finite-size
scaling of the dimer order parameter $M_{L}$ using the ansatz
\begin{equation}
M_{L}(D) \simeq L^{-\beta/\nu} f_M\big(|D-D_c|L^{1/\nu}, L^{-\omega}\big),
\end{equation}
where $\beta$ and $\nu$ are the critical exponents for the order parameter and
correlation length, respectively, and $f_M$ is a nonuniversal function. Critical
exponents are extracted by adjusting $\mu_1$ and $\mu_2$ so that $M_{L} L^{\mu_1}$
versus $D$ intersects and $M_{L} L^{\mu_1}$ versus
$|D-D_c|L^{\mu_2}$ collapses for all $L$; then $\beta = \mu_1/\mu_2$
and $\nu = 1/\mu_2$. The finite-size scaling results for $M_L$ at
$\vartheta/\pi = -0.4950$ with $L=128,192,256$ are shown in
Supplementary Fig.~\ref{fig:SMfss}. From the scaling ansatz we obtain
$D_c \approx -0.6464$, $\mu_1 = \beta/\nu \approx 0.126$, and
$\mu_2 = 1/\nu \approx 1.02$, yielding $\beta \approx 0.124$ and $\nu \approx 0.98$,
in full agreement with the Ising transition.

\begin{figure}[htbp]
\centering
\includegraphics[width=0.50\columnwidth, clip]{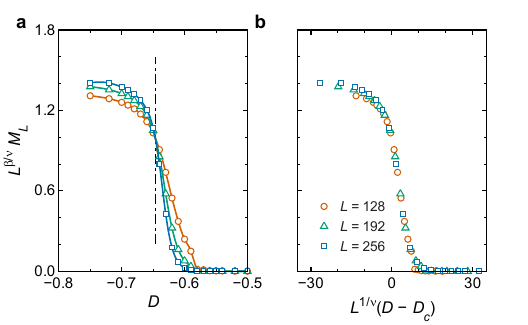}
\renewcommand{\figurename}{\textbf{Supplementary Fig.}}
\caption{\textbf{Finite-size scaling of the dimer order parameter.} Dimer order
parameter $M_{L}$ as a function of $D$ in
the Kitaev--$\Gamma$ spin chain with SIA at $\vartheta/\pi = -0.4950$. The chain
lengths $L$ chosen are 128 (red circle), 192 (green triangle), and 256 (blue
square).}
\label{fig:SMfss}
\end{figure}

\smnote{Order-parameter jumps at $D=-0.7$}{SM:D070}

This section collects the data underlying the statement in the main text that the
dimer--Haldane transition is of first order at $D=-0.7$.

In the Kitaev spin chain with SIA, the ground state is dimerized for
$D<D_c=-0.6551(2)$~\cite{SMLuo2023PRB}. Hence, at $D=-0.7$ and the Kitaev-only
coupling angle $\vartheta=-\pi/2$, the system resides in the dimerized phase.
Turning on the $\Gamma$ interaction (i.e., increasing $\vartheta$ above $-\pi/2$)
progressively suppresses this dimerization. Supplementary Fig.~\ref{fig:SMd070}a displays
the dimer order parameter $M_L$ for $L=128$, $192$, and $256$ at $D=-0.7$. Deep in
the dimerized regime, $M_L$ is robust and exhibits only weak finite-size effects.
As $\vartheta$ moves further away from the Kitaev point, $M_L$ first decreases
slightly and then drops sharply to zero at $\vartheta_c/\pi=-0.4909(2)$. The
string order parameter $\mathcal{O}_H$ also changes abruptly across the transition
[Supplementary Fig.~\ref{fig:SMd070}b]. These discontinuities unambiguously identify the
dimer--Haldane transition as first order at this $D$.

\begin{figure}[htbp]
\centering
\includegraphics[width=0.50\columnwidth, clip]{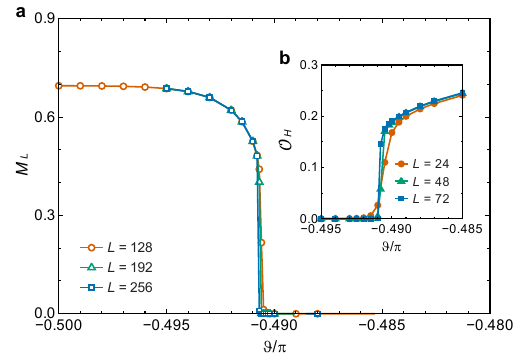}
\renewcommand{\figurename}{\textbf{Supplementary Fig.}}
\caption{\textbf{Order-parameter jumps at the first-order dimer--Haldane
transition.} \textbf{a}, Dimer order parameter $M_{L}$ as a
function of $\vartheta$ in the Kitaev--$\Gamma$ spin chain with SIA $D = -0.7$
for different chain lengths $L=128$ (red circles), $192$ (green triangles), and
$256$ (blue squares) under OBC. \textbf{b}, String order parameter
$\mathcal{O}_{H}$ as a function of $\vartheta$ for different chain lengths
$L=24$ (red circles), $48$ (green triangles), and $72$ (blue squares) under PBC.}
\label{fig:SMd070}
\end{figure}


\end{document}